\documentclass[twocolumn]{aa}
\usepackage{graphicx}
\usepackage{txfonts}
\usepackage{natbib}
\bibpunct{(}{)}{;}{a}{}{,}
\newcommand{\HI}{\mbox{H {\sc i}}}
\newcommand{\HII}{\mbox{H {\sc ii}}}
\begin{document}
   \title{Extra-planar \HI\ in the starburst galaxy NGC 253}

   \subtitle{}

   \author{R. Boomsma\inst{1}\and
          T.A. Oosterloo\inst{2}\and
	  F. Fraternali\inst{3,2}\and
	  J.M. van der Hulst\inst{1}\and
	  R. Sancisi\inst{4,1}
          }

   \offprints{R. Boomsma}

   \institute{Kapteyn Astronomical Institute, P.O. Box 800, 9700 AV Groningen\\
              \email{boomsma@astro.rug.nl}
		\and
		ASTRON, Dwingeloo
		\and
		Theoretical Physics, University of Oxford
		\and
		INAF-Osservatorio Astronomico di Bologna, Bologna
             }

   \date{Received July 22, 2004, accepted September 23, 2004}

   \abstract{

Observations of the nearby starburst galaxy \mbox{NGC 253} in the
21-cm line reveal the presence of neutral hydrogen in the halo, up to
12 kpc from the galactic plane. This extra-planar \HI\ is found only
in one half of the galaxy and is concentrated in a half-ring
structure and plumes which are lagging in rotation with respect to the
disk. The \HI\ plumes are seen bordering the bright H$\alpha$ and
X-ray halo emission.  It is likely that, as proposed earlier for the
H$\alpha$ and the X-rays, also the origin of the extra-planar \HI\ is
related to the central starburst and to the active star formation in
the disk. A minor merger and gas accretion are also discussed as
possible explanations.\\
 The \HI\ disk is less extended than the
stellar disk.  This may be the result of ionization of its outer parts
or, alternatively, of tidal or ram pressure stripping.
   

   }

   \maketitle
%

\section{Introduction}

\mbox{NGC 253} is a barred Sc galaxy in the Sculptor group of
galaxies, at a distance of 3.9 Mpc \citep{kar03}. It is one of
best nearby examples of a nuclear starburst galaxy. Outside of its
very active nuclear region, \mbox{NGC 253} shows a global morphology
in gas and stars expected for a normal spiral galaxy.\\
\begin{figure*}[!ht]
\includegraphics[width=\textwidth]{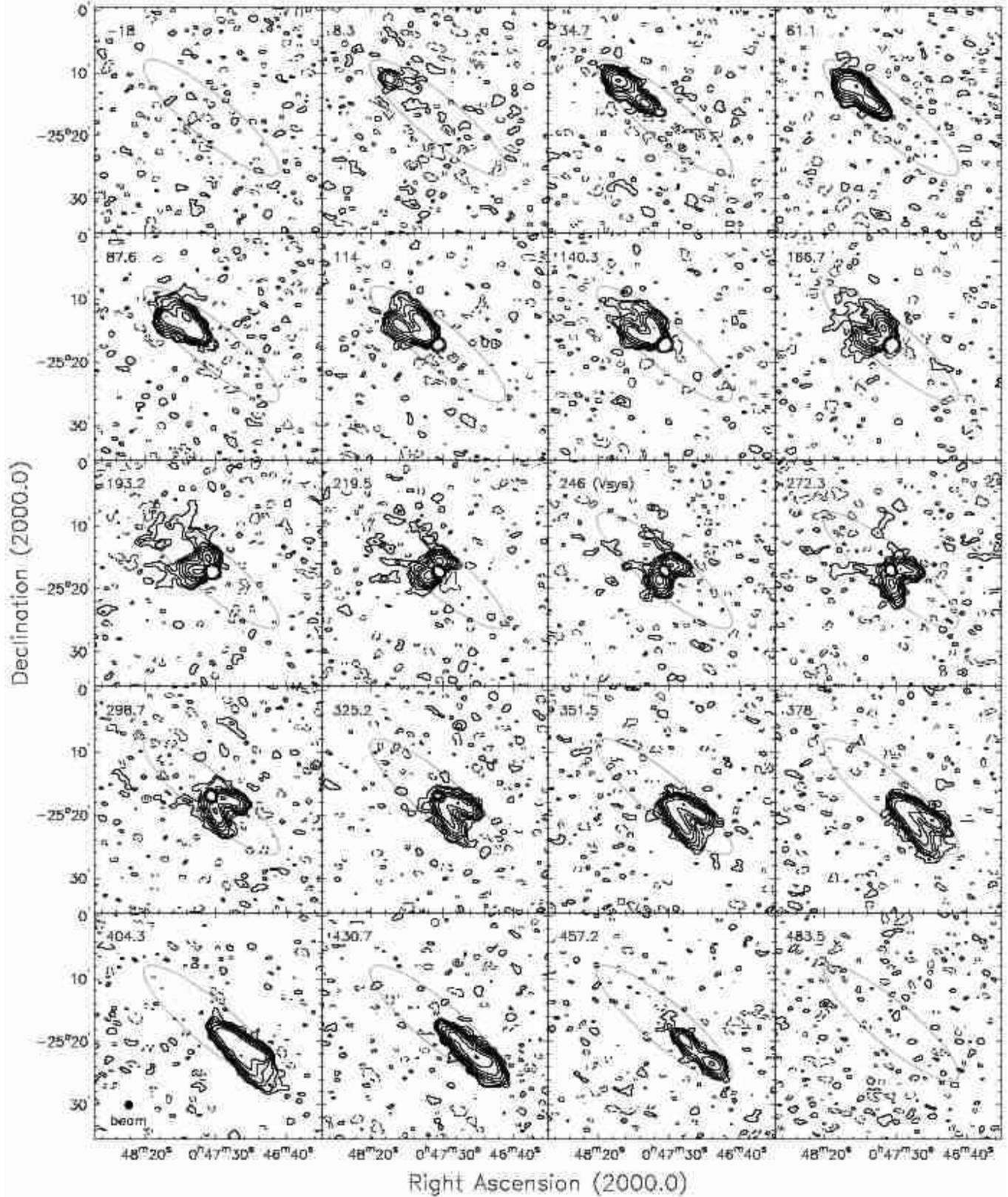}
\caption{NGC 253: \HI\ channel maps. The ellipse outlines the bright optical disk (R$_{25}$). Contours are at -4.6, -2.3, 2.3 ($1.75\sigma$), 4.6, 9.2, 18.4, 36.8, 73.6, 147.2, and 294.4 mJy/beam, with negative contours dashed. The heliocentric radial velocities (km/s) are given in the upper left corners. The 70'' beam is shown by the filled circle in the bottom left panel.}
\label{channels}
\end{figure*}
The nuclear starburst is thought to be forming stars at a fairly
high rate and to produce a superwind \citep{hec90} which brings
material into the halo. Starburst, disk and halo of \mbox{NGC 253}
have been studied at various wavelengths. The X-ray emission has been
observed with ROSAT \citep{pie00}, with Chandra \citep{str02} and with
XMM-Newton \citep[][ data on the halo unpublished]{pie01}.  The ROSAT
observations have revealed diffuse soft X-ray emission from the
nucleus, the disk and a halo extending up to a distance of 9 kpc from
the disk. Narrow band H$\alpha$ imaging \citep{hoo96} has shown the
presence of a diffuse ionized gas (DIG) component. The radio continuum
study by \citet{car92} has revealed a bright disk and a radio halo
extending to a height of at least 9 kpc above the plane of \mbox{NGC
253}. Infrared studies with IRAS \citep{ric93,alt98} and ISO
\citep{rad01} show that there is also dust in the halo region.\\
Studies of a number of nearby spirals with active star formation
show considerable amounts of \HI\ in their halos
\citep{fra01,fra03}. Also the well-known IVCs and HVCs of our Galaxy
\citep{wak97,woe04} may represent a similar halo component. The origin
of such cold halo gas is not known with certainty, but it is thought
that galactic fountains set up by the active star forming regions of
the disk may be at least partly responsible.  One would
therefore expect a significant amount of extra-planar gas also in
\mbox{NGC 253}, because of the starburst and of the star formation
which is very active over a large fraction of its disk. But so far no
extra-planar \HI\ had been found in this galaxy
\citep[cf.][]{puc91,kor95}.\\ Here we present the results of a new and
deeper \HI\ study of \mbox{NGC 253} which shows that there is indeed
\HI\ in the halo of this starburst galaxy. Although the central
starburst and the active star formation in the disk are likely to
cause gas to flow into the halo, there can be other explanations for
the origin of this extra-planar \HI. For example, in many galaxies,
including the Milky Way, observations show ongoing accretion. We will
discuss the different possibilities for the origin of this
extra-planar \HI.\\
\begin{figure*}[!ht]
\includegraphics[width=\textwidth]{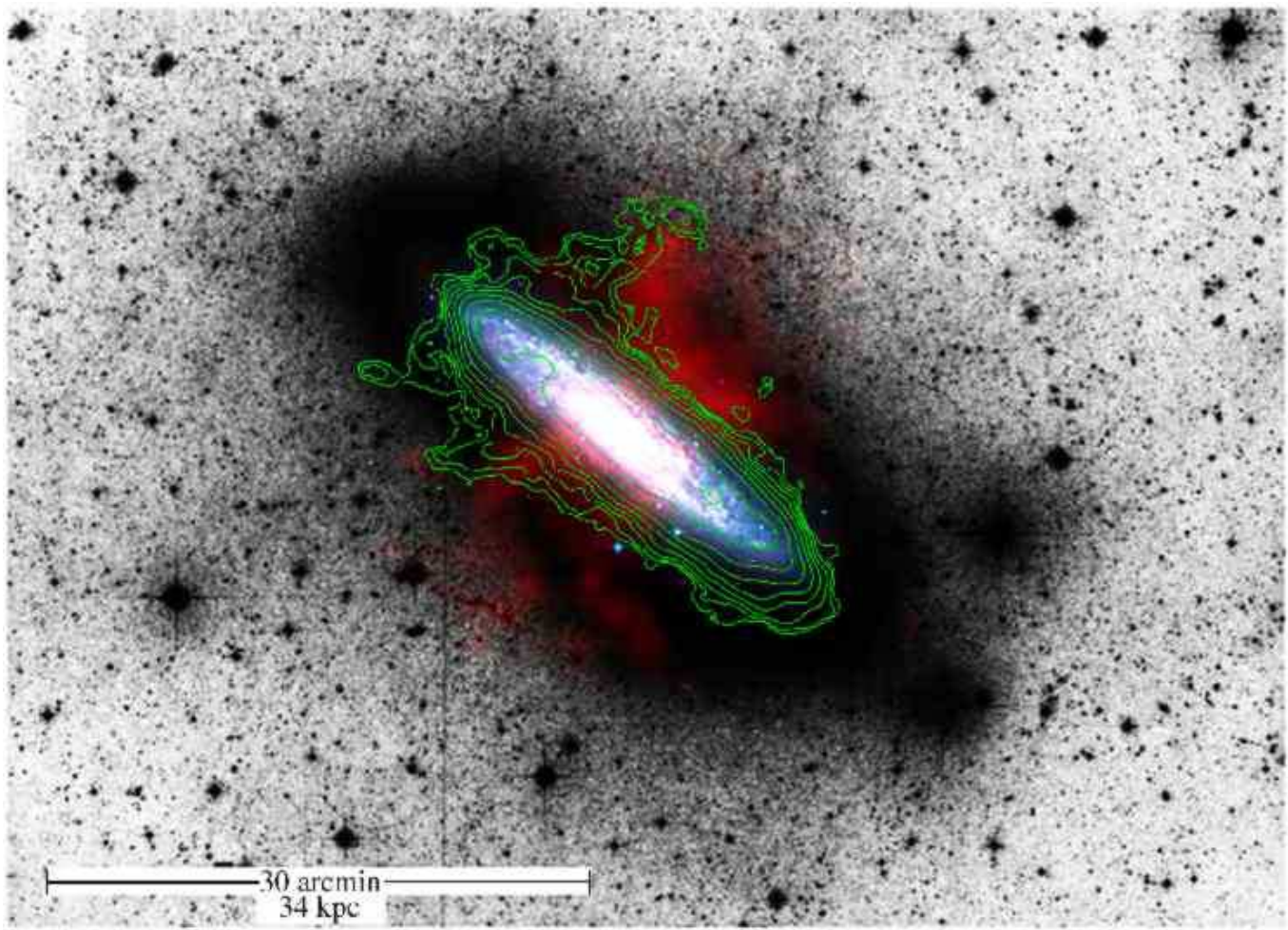}
\caption{Multi-wavelength image of the nearby spiral galaxy \mbox{NGC
253}. The deep optical image is from \citet{mal97}, the blue shows the
DSS optical disk, the red the X-ray emission (0.1 - 0.4 keV) from
ROSAT \citep{pie00}, and the green contours our Compact Array \HI\
observations. Contours are at 0.18, 0.36, 0.72, 1.4, 2.9, 5.8, 12 and
23$\times 10^{20}$ cm$^{-2}$. The circular beam has a full half-width of
70$^{\prime\prime}$.\label{fig1}}
\end{figure*}

\section{Observations and reduction}

We have obtained new 21-cm line observations with the Australia
Telescope Compact Array (ATCA) and have combined these with archival
\HI\ data obtained by \citet{kor95}. These new observations have been
taken between February and September 2002 in four different
configurations (1.5A, 1.5B, 750B, EW367). The total bandwidth is 8 MHz
with 512 spectral line channels. The total integration time,
including the data of \citet{kor95}, is $8\times12$ hours.\\ The data
have been reduced using the MIRIAD software package \citep{sau95}. The
continuum has been derived from the line-free channels and subtracted
in the {\it uv}-plane. Subsequently, the data have been
Fourier-transformed using robust weighting 0 \citep{bri95},
Hanning smoothed, and CLEANed \citep{hog74,cla80}. The final dataset
has been obtained by smoothing to a circular beam of
70$^{\prime\prime}$. The resulting r.m.s. noise is 1.4 mJy/beam
($0.17$ K) and the velocity resolution is 13.2 km s$^{-1}$. We
tried further smoothing, but that does not reveal any new
\HI\ features. For the subsequent analysis of the data we have used
the GIPSY package \citep{vdh92,vog01}. The channel maps, after
averaging in groups of three, are shown in Fig. \ref{channels}.

\section{Results}

\subsection{The extra-planar \HI}

\begin{figure}[!ht]
\includegraphics[width=\columnwidth]{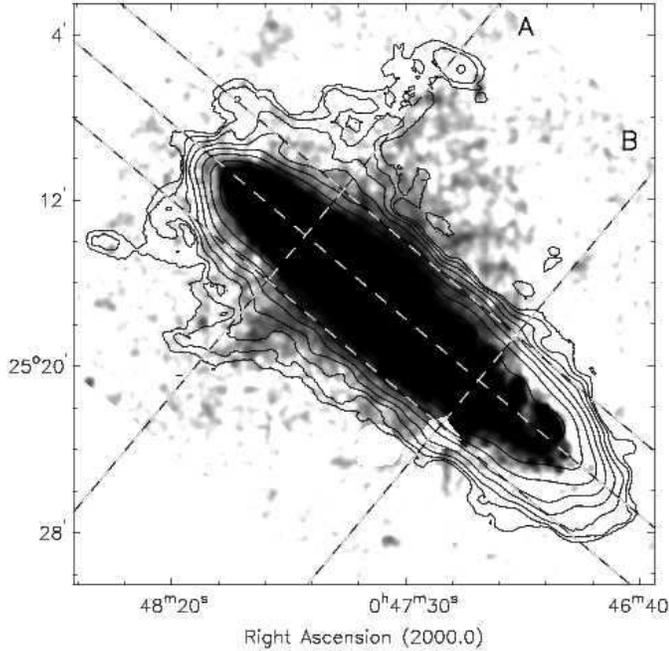}
\caption{\HI\ contours overlaid on an H$\alpha$ image (smoothed with a
Gaussian beam of 20$^{\prime\prime}$) of \mbox{NGC 253}
\citep{hoo96}. Contours are the same as in \mbox{Fig. \ref{fig1}}. The
dashed lines parallel to the major axis indicate the locations of the
slices shown in Fig. \ref{fig3}. These are separated by
$3^{\prime}$. The lines labeled 'A' and 'B' correspond to the slices
in Fig. \ref{fig5}, and are at $5.5^{\prime}$ distance from the
minor axis}.\label{fig2}
\end{figure}

\begin{figure}[!ht]
\includegraphics[width=\columnwidth]{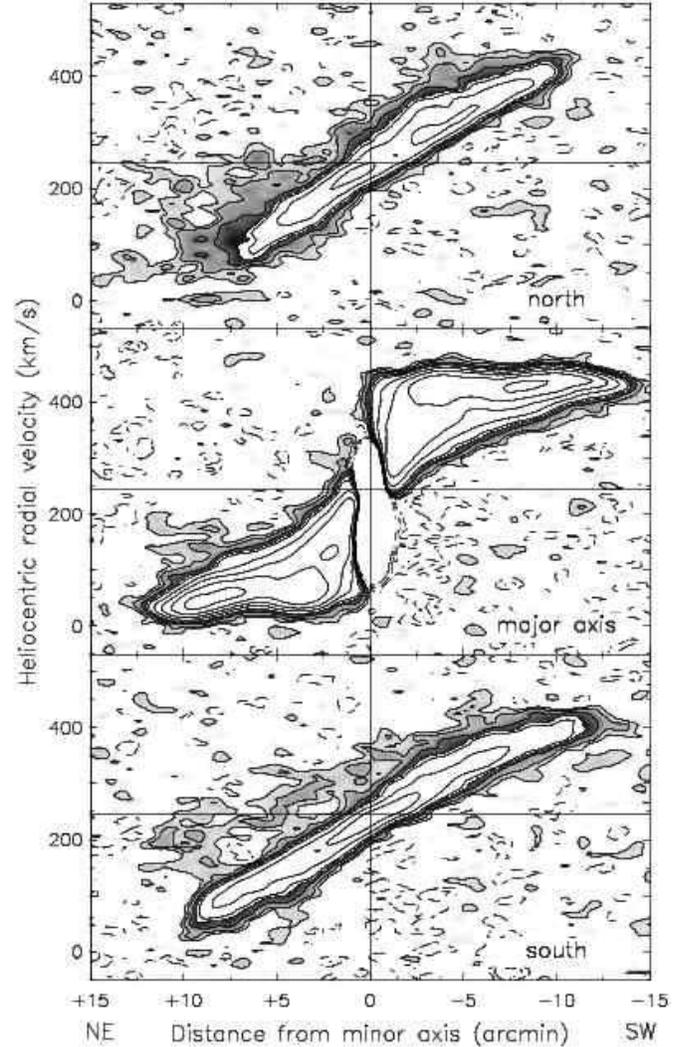}
\caption{Position-velocity diagrams along the major axis of 
NGC 253 (middle panel) and parallel to the major axis to the NW (top) 
and to SE (bottom).The positions of the slices with respect to the galaxy 
are shown in Fig. \ref{fig2}. The horizontal lines indicate the systemic velocity. Contours are at -21.6, -10.8, -5.4, 5.4 ($1.5\sigma$), 10.8, 21.6, 43.2, 86.4, 192.8, 385.6 and 770.2 mJy/beam. Negative contours are dashed.\label{fig3}}
\end{figure}

\begin{figure}[!ht]
\includegraphics[width=\columnwidth]{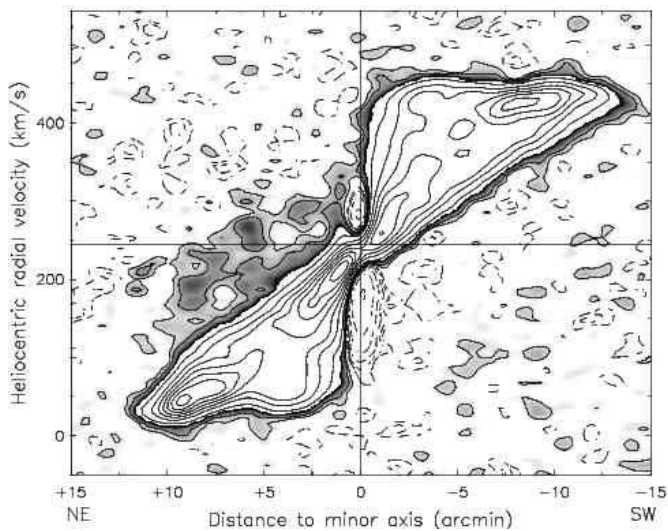}
\caption{Overall \HI\ position-velocity map of \mbox{NGC 253}. All the
emission has been integrated in the direction parallel to the minor
axis. The halo emission is visible at the northeast side of the
galaxy, near the systemic velocity. The hole in the middle is due to
\HI\ absorption against the bright radio continuum source in the
nucleus. Contours are at -4, -2, -1, -.5 -.25 -.125, .125, .25, .5, 1,
2, 3, 4, 5, 6 and 7 Jy/beam. Clearly, the anomalous gas is only
present on the northeast side of the galaxy.}
\label{fig4}
\end{figure}

\begin{figure}[!t]
\includegraphics[width=\columnwidth]{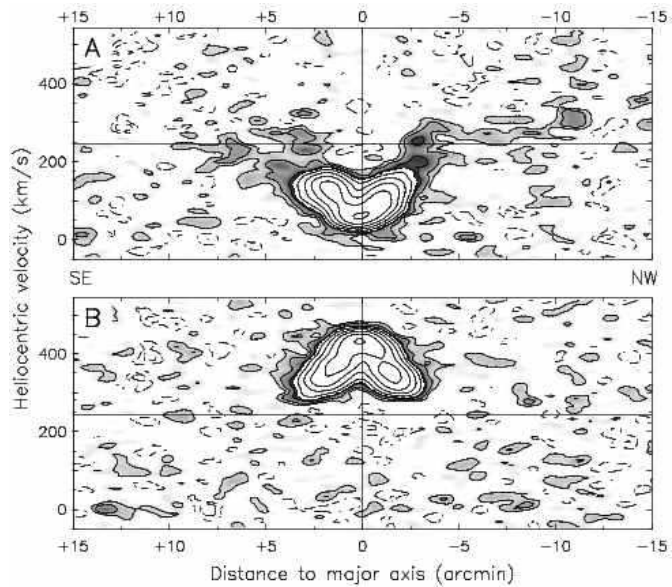}
\caption{Slices parallel to the minor axis from southeast (left) to
northwest. Top panel is northeast of the centre; bottom panel
southwest (see for positions on the sky Fig. \ref{fig2}). The emission
has been integrated over 2$^{\prime}$ wide strips. The horizontal line
indicates the systemic velocity of \mbox{NGC 253} (\mbox{245 km
s$^{-1}$}). Contours are at -.06 -.03, .03, .06, .12, .24, .48, .96,
1.92 and 2.84 Jy/beam. Note the extended emission near systemic
velocity, that is visible in the upper panel, but absent in the lower
one.}
\label{fig5}
\end{figure}

\mbox{Figure \ref{fig1}} shows the \HI\ distribution together with the
DSS optical and the ROSAT soft (0.1 - 0.4 keV) X-ray images. The \HI\
contour map shows two components: the \HI\ disk and, in the northeast
half of the galaxy, gas concentrations in the halo region with a plume
extending 11$^{\prime}$ from the major axis to the NW, or 12 kpc for
the assumed distance of 3.9 Mpc. A smaller plume reaches a distance of
8$^{\prime}$ from the major axis to the SE. These \HI\ plumes border
the X-ray halo emission \citep{pie00} and the H$\alpha$
(see Fig. \ref{fig2}) \citep{hoo96} at their northern side. The
spatial relationship between the different components suggests a
common origin. This is further discussed in section 4.1.\\
The \HI\ kinematics in \mbox{NGC 253} is illustrated in
Figs. \ref{channels}, \ref{fig3}, \ref{fig4} and \ref{fig5}. Figure
\ref{fig3} shows three position-velocity diagrams, one along the major
axis, and two parallel to and on either side of the major axis (see
Fig. \ref{fig2} for the positions of the cuts on the sky). The \HI\
disk shows the regular pattern of differential rotation with an
approximately flat rotation curve. In the plot along the major axis,
the hole in the direction of the galaxy centre is due to \HI\
absorption against the bright radio continuum source. In addition to
the bright, normal, rotating \HI\ disk there is another component
which shows up as weak emission close to systemic velocity. This is
seen only on the approaching NE side of the galaxy. In the major axis
map this weak \HI\ emission is only present in the outer parts,
between 8 and 12 arcmin, whereas in the two slices away from the major
axis it is seen at all radii and also at ``forbidden''
velocities. In these two cuts, similar weak \HI\ emission seems to be
present also in the receding SW part of the galaxy at the high
rotation velocity side. This may, however, partly be the result of
beam-smearing from parts of the galaxy nearer to the major axis, where
such velocities are forthcoming.\\
Figure \ref{fig4} shows the overall space-velocity distribution of
\HI\ in \mbox{NGC 253} in the direction parallel to the major axis. It
has been obtained by integrating in the direction of the minor axis.
The presence and extent of the weak emission (to be called henceforth
the "anomalous \HI") close to systemic velocity and on the NE half of
the galaxy are particularly clear in this map. Part of this gas has
forbidden velocities. In the SW half of \mbox{NGC 253} there is no
indication at all of such a component. The presence of the
anomalous \HI\ can easily be seen in the channel maps (Fig. \ref{channels})
in the velocity range from 166.7 to 298.7 km/s. The kinematics of
this anomalous \HI\ is similar to that of the \HI\ halos of \mbox{NGC
891} \citep{swa97, fra03} and of \mbox{NGC 2403} \citep{sch00, fra01}
which rotate more slowly than the gas in the disk and, in
position-velocity diagrams, also show up as emission close to the
systemic velocity. However, contrary to what has been found in
\mbox{NGC 891} and \mbox{NGC 2403}, the distribution of the anomalous
gas in \mbox{NGC 253} presents an intriguing asymmetry between the two
halves of the galaxy.\\ Figure \ref{fig5} shows position-velocity
diagrams for two slice positions (cf. Fig. \ref{fig2}) parallel to the
minor axis of \mbox{NGC 253}.  The top panel is for the northeast half
of the galaxy, the bottom one for the southwest half. In the top
panel, the anomalous \HI\ shows up as low intensity emission at
approximately systemic velocity. It appears to be smoothly connected
to the thin disk rather than being a kinematically distinct component.
In the bottom panel there is no trace at all of such anomalous \HI\ near the systemic velocity, in agreement with Figs. \ref{fig3} and \ref{fig4}.\\

\begin{figure}[!t]
\includegraphics[width=\columnwidth]{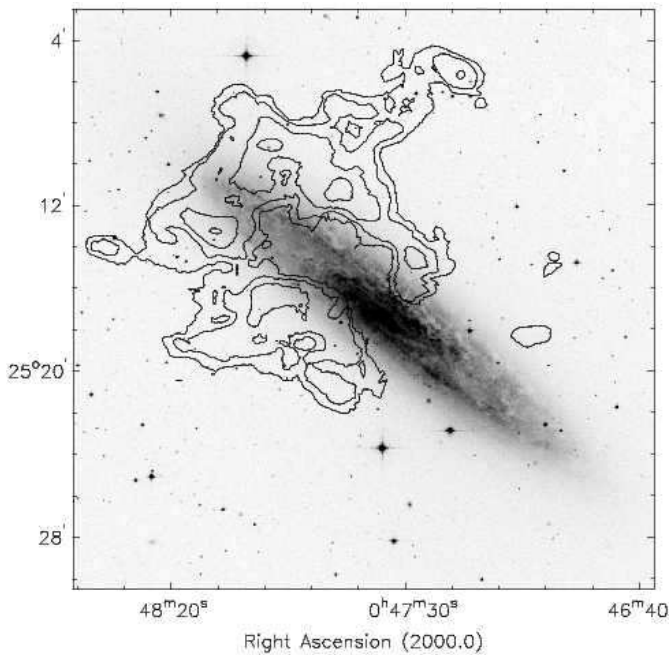}
\caption{Contour map of the anomalous \HI\ superimposed on
the optical image of \mbox{NGC 253} from the Digitized Sky
Survey. Contours are at 1.8, 3.6, 7.2 and 14$\times 10^{19}$
cm$^{-2}$.}
\label{fig7}
\end{figure}

We have separated the anomalous \HI\ from the normal, rotating \HI\
disk and constructed a map of its distribution on the sky
(Fig. \ref{fig7}). The separation has been obtained by masking out the
emission from the regular, rotating disk. The masking has been done by
visual inspection of the position-velocity diagrams parallel to the
major axis as those shown in Fig. \ref{fig3}.\\ The total mass of the
anomalous \HI\ is $8\times10^{7}$ M$_{\odot}$. This is about 3\% of
the total amount of \HI\ in \mbox{NGC 253} ($=2.5\times10^{9}$
M$_{\odot}$). The column densities in the plumes reach peak values of
about $5\times10^{19}$ cm$^{-2}$. It should be noted, however, that
beam smearing is likely to be important and locally the densities may
be considerably higher.\\ This anomalous \HI\ is found only in the
northeast half of \mbox{NGC 253} and is all concentrated at the border
of the bright optical disk. This particular aspect of the distribution
of the anomalous gas with respect to the optical and \HI\ disk -
i.e. its absence in the inner parts of the galaxy - can be directly
veryfied on Fig. \ref{fig3}, where the middle panel shows that the
anomalous \HI\ is lacking inside the inner 8 arcmin. Over the whole
southwest half of the galaxy there is no detection at all. The 3-D
distribution of the anomalous gas is discussed further in section
4.1.\\ However, this intriguing northeast/southwest asymmetry and the
absence of signal in the direction of the bright optical disk should
be taken with caution as the derived signal is close to the detection
limit. It needs to be confirmed with new, more sensitive
measurements.\\

\subsection{The small \HI\ disk}

A striking characteristic of the \HI\ disk is its small size as
compared to the optical. Normally, the \HI\ disks of spiral galaxies
are larger than the stellar disks \citep[see
e.g.][]{bos81,bro92,ver97}. For \mbox{NGC 253} the \HI\ diameter at
the 1 M$_{\odot}$ pc$^{-2}$ contour level ($1.25\times10^{20}$
cm$^{-2}$) is 26$^{\prime}$, whereas the optical diameter D$_{25}$ is
$27^{\prime}.7$ \citep{pen80}. The deep optical exposure by
\citet{mal97} in Fig. \ref{fig1} shows that the stellar disk reaches
even further out. The \HI\ disk on the contrary shows a
cutoff. Although our map goes deeper than that of \citet{kor95}, we
find only a slightly larger radial extent.\\ Figure \ref{fig4} shows
that the \HI\ disk of \mbox{NGC 253} is asymmetric, in extent as well
as in the kinematics. On the northeast side, where also the anomalous
\HI\ is seen, the disk appears to be less extended. The channel
maps in Fig. \ref{channels} show a structure and a change of position
angle with radius suggesting an outer warping of the \HI\ disk \citep[see
also tilted ring analysis by][]{puc91}.\\ A lopsidedness and a
minor warp as found in this starburst galaxy are, however, common
features of spiral galaxies. Apart from its relatively small \HI\
radius as compared to the optical, \mbox{NGC 253} looks normal.  The
overall structure and kinematics of its \HI\ disk are regular
and unperturbed. They do not show large disruptions (even in the central
parts and near the starburst) or other anomalies that might point to
recent, major merger events.\\

\section{Discussion}

The present \HI\ observations of \mbox{NGC 253} have revealed the
presence of extra-planar gas in the northeast half of the
galaxy. Furthermore, they have shown that the \HI\ disk is
unperturbed and apparently truncated well inside the stellar disk.\\
These results will be discussed here in connection with the nuclear
starburst and galactic fountains in the central regions of the disk.
Also the possibility of a minor merger and gas infall will be given
some consideration.\\
\subsection{Structure and kinematics of the extra-planar \HI}

We have constructed models of the 3-D structure and kinematics of
\mbox{NGC 253} consisting of a thin \HI\ disk and a thick disk with
different kinematics. The picture we derive for the latter is that of
a thick disk with a large, central cavity. As seen in Figs. \ref{fig3} and \ref{fig4},
this structure, resembling that of a half-ring, seems only
present in the northeast side of the galaxy. The neutral gas forming
it is located above and below the outer border of the northeast half
of the disk. The ring rotates more slowly than the thin disk. Radial
motions, both inward and outward, may also be present. The \HI\ plumes
seen above and below the disk may be filaments or wall-like extensions
of the ring to high z-distances. The cavity in the halo outlined by
ring and plumes has a radius of about 6 kpc.\\

Two questions arise in connection with the puzzling
northeast/southwest halo asymmetry.  One is whether there is no
extra-planar gas at all on the southwest part of \mbox{NGC 253},
either in neutral or ionized form.  The other is what might have
caused such asymmetry.\\ As to the first question, it should be noted,
as already pointed out earlier, that in the northeast part of the
galaxy the \HI\ emission is barely detected and a factor 2 weaker
emission in the southwest would be below detection. Therefore deeper
\HI\ observations are necessary to confirm the asymmetry and place
better upper limits on the \HI\ emission in the southwest halo region.
At any rate, the present limits already imply an unexpectedly low
column density of \HI.  For comparison, \HI\ densities as found in the
halo of \mbox{NGC 891} and of \mbox{NGC 2403} would have been detected
with the present observations of \mbox{NGC 253}.

 There is the possibility that the asymmetry in \HI\ is simply caused
by complete ionization over most of the halo including all of the
southern part. This will be briefly considered below.\\ As to the
cause of the asymmetry, it is interesting to note that an asymmetry is
also seen in the X-rays plumes and in the H$\alpha$, although not as
pronounced as in \HI, whereas no striking asymmetry is present in the
inner galaxy disk either in the optical, in the infrared (see the Two
Micron All Sky Survey, Jarrett et al. 2003, and the 60 $\mu$m dust,
Rice 1993), or in the \HI. In H$\alpha$ and the radio continuum
\citep{car92} the northern spiral arm is brighter than the southern
arm. It is plausible that these asymmetries are related to the
asymmetry found for the extra-planar \HI\ emission as will be
discussed in section 4.2.

\subsection{Galactic fountains and starburst}

\begin{figure}[!ht]
\includegraphics[width=\columnwidth]{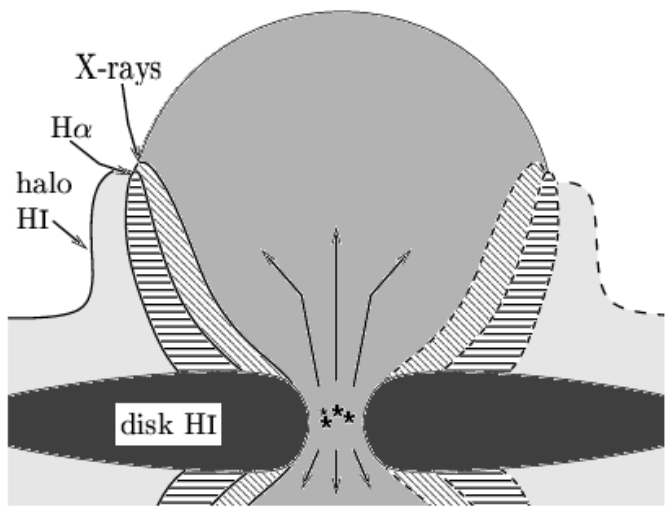}
\caption{Same as figure 11d of \citet{str02}, but with an \HI\
shell added. The various components are
described in the text. The dashed lines on the right outline parts that
have not been detected.}
\label{fig8}
\end{figure}

The main source of energy input into the halo is probably the central
starburst. A significant amount of the flux comes, however, also from
the region of active star formation in the central regions of the
disk, within a 4 kpc radius \citep{ulv00}, which is particularly
bright in the radio continuum \citep[cf.][]{car92}, the H$\alpha$ and
the soft X-rays.  The superbubble expanding into the halo is therefore
the combined effect of the nuclear starburst and to a large extent
also of the {\bf star formation} in the inner disk or ring.\\ The
relative locations of \HI, radio continuum, X-rays and H$\alpha$
provide important clues for understanding the 3-dimensional picture.
There is a correspondence between \HI\ and radio continuum
features. The shapes of the halo extensions in the 0.33 GHz image
\citep{car92} look similar to the shape of the \HI\ halo
distribution. The \HI\ plume (to the SE) appears at the same position
as the radio continuum spur observed by \citet{car92}. This spur might
mark the eastern edge of the outflow cone of matter.  As already
noted, the H$\alpha$ and the X-ray emission are displaced with respect
to the \HI\ plumes. Similarly, there also seems to be a displacement
between the X-rays and the H$\alpha$ emission \citep[see Fig. 10b
of][]{str02}.  \citet{str02} have made several models to explain the
X-ray and the H$\alpha$ observations. The shape of the \HI\
distribution as derived here, supports their model 5 (see their
Fig. 11d and Fig. \ref{fig8} in this paper). In this model, the X-ray
emission originates from a shock-heated shell around a tenuous hot
superbubble, which is surrounded by a cooler shell of \HII\ and \HI\
that was dragged up from a thick disk. The cartoon in Fig. \ref{fig8},
partly adapted from \citet{str02}, illustrates schematically the gas
structure of \mbox{NGC 253}. The thick layer in this picture is formed
by the \HI\ clouds in the lower halo. The cavity in the halo cloud
layer is filled by a hot (about $10^6$ K) halo medium that is observed
in the soft X-rays \citep[see sketch (Fig. 10) in ][]{pie00}.\\

In the picture suggested here, the absence of extra-planar \HI\ over
the largest part of the halo region of \mbox{NGC 253} (the \HI\
cavity described above), especially when compared with other galaxies
with \HI\ halos, raises interesting questions. Why is there less halo
gas in \mbox{NGC 253} than in \mbox{NGC 891} and 2403? and even more
striking, if the halos of \mbox{NGC 891} and 2403 are the result of
galactic fountains, why should there be less gas in the halo of
\mbox{NGC 253}, where there is a starburst and a high degree of star
formation activity?  Is it possible that there was a neutral gas halo
in \mbox{NGC 253} and that it has been swept away by the expanding
superbubble? Or, alternatively, could there be extra-planar gas
everywhere, filling the halo of \mbox{NGC 253} also in the southwest
part, and be all ionized?  This should be addressed with deeper
H$\alpha$ studies, although it seems to be contradicted by the
presence of fainter H$\alpha$ emission in the SW part where the \HI\
is missing. The extraplanar \HI\ in the northeast side might be part
of the gas swept away, or, in the case of a fully ionized halo, it
could be cold (or cooling down) gas not ionized because of its higher
density.\\

\subsection{Minor merger, gas infall}

 The observed extraplanar \HI\ could be accreted gas from a merger
with a satellite. Perhaps, such a minor merger could also explain the
triggering of the starburst in an otherwise normal, non-interacting
galaxy like NGC 253. The age of the starburst and the superbubbles is
a few times 10$^{7}$ yr. It would take approximately 10$^{9}$ yr to
spread accreted gas symmetrically over the whole disk. These
timescales for the infall and for the starburst would all be
consistent with the observed asymmetric \HI\ halo distribution. There
are also indications from the optical (see section 4.4) that a minor
merger may have occurred.\\ A minor merger would naturally account for
the anomalous \HI\ and the asymmetric picture. The question, however,
is whether it could produce the remarkable velocity-space continuity
and also the symmetry with respect to the major axis of the
extra-planar \HI\ with the disk of \mbox{NGC 253} pointed out above
and shown in Fig. \ref{fig5}. The X-rays and H$\alpha$ would still
have to be explained by the superwind from the starburst, while their
apparent connection with the \HI\ plumes could be due to the superwind
hitting the tidal wreckage. All considered, a merger event seems
unlikely in this case.

\subsection{The truncated \HI\ disk}

 What causes the cutoff at the edge of the \HI\ disk? According to
\citet{bla97} there is H$\alpha$ emission along the major axis of
\mbox{NGC 253} beyond the SW edge of the \HI\ disk and its
velocity is in agreement with that of the \HI. They conclude that most
likely the hot stars in the disk of \mbox{NGC 253} itself are
responsible for the ionization. For the disk to be ionized in its
outer parts by the stars and the starburst, its outer layer has to be
warped.  As noted in the previous section, there is evidence in the
present data for the presence of such a warp as also found by
\citet{puc91} and Koribalski et al.(1995). The position and
inclination angles of the disk change somewhat outside a 9'
radius.  Ionization of the outer hydrogen disk by the central
starburst seems, therefore, to be possible. Perhaps one should also
consider the possibility of strong UV radiation from the halo region.
In this connection, the reason why the extra-planar \HI\ on the
northeast side is not ionized may lie in its relatively high
densities.\\ A different mechanism for reducing the size of the \HI\
disk in \mbox{NGC 253} is ram pressure stripping. However, the
Sculptor group, of which \mbox{NGC 253} is a member, is not a compact
group but a loose filament of galaxies extended along the line of
sight \citep{jer98, kar03}.  It is therefore unlikely to contain a
significant amount of intra-group medium capable of causing the
stripping.\\ A third possibility is that the gaseous disk of \mbox{NGC
253} has been truncated as the result of a minor merger. A number of
observational elements suggest that \mbox{NGC 253} might have
undergone an encounter. The galaxy shows an unusual stellar halo
\citep{mal97} with an optical loop extending to the south at the
southwest edge of the galaxy \citep{bec82} (see also
Fig. \ref{fig1}). A minor merger has already been mentioned above as a
possible, but unlikely cause for the anomalous gas.

\section{Summary}

Extra-planar \HI\ concentrations have been observed in the northeast
half of the starburst galaxy \mbox{NGC 253}. Plumes of \HI\ are seen
reaching as high as 12 kpc above the disk. They seem to border the
H$\alpha$ and X-ray superbubble on its northern side.  Apparently,
this extra-planar \HI\ has an half-ring structure and, similar to the
\HI\ halos found in \mbox{NGC 891} and 2403, is rotating more slowly
than the gas in the disk. The cavity in the \HI\ halo and in
particular the apparent absence of extraplanar \HI\ in the southwest
half of \mbox{NGC 253} are puzzling results.  More sensitive \HI\
observations are necessary to confirm them.  It is likely that, if
real, they are due to the combined effects of the central starburst
and star formation in the disk.\\

 The \HI\ disk is truncated. Some explanations have been mentioned. 
A possibility is that the warped outer parts have been ionized 
by the photons from the starburst and the active star formation 
in the disk. However, ram pressure or tidal stripping cannot be 
ruled out.\\

\begin{acknowledgements}
We thank Jay Gallagher and Jacqueline van Gorkom for stimulating
discussions and Albert Bosma for his valuable criticism on the
manuscript. We are grateful to Charles Hoopes for providing the X-ray
and the H$\alpha$ images. The Australia Telescope is funded by the
Commonwealth of Australia for operation as a National Facility managed
by CSIRO.\\
\end{acknowledgements}

\end{document}